\documentclass[twocolumn,showpacs,aps,prl]{revtex4}
\usepackage[dvips]{graphicx}

\usepackage{amsmath}

\begin{document}

\title{ Warburg's impedance revisited}
\author{G. Barbero}
\affiliation{ Dipartimento di Scienza Applicata del Politecnico, Corso Duca degli
Abruzzi 24, 10129 Torino, Italy.\\
{\rm and}\\
National Research Nuclear University MEPhI (Moscow Engineering Physics Institute), Kashirskoye shosse 31, 115409 Moscow, Russian Federation.\\
}
\date{\today}
\begin{abstract}
The derivation of Warburg's impedance presented in several books and scientific papers is reconsidered. It has been obtained by assuming that the total electric current across the sample is just due to the diffusion, and that the external potential applied to the electrode is responsible for an increase of the bulk density of charge described by Nernst's model. We show that these assumptions are not correct, and hence the proposed derivations questionable.  A correct  determination of the electrochemical impedance of a cell of an insulating material where are injected external charges of a given sign, when the diffusion and the displacement currents are taken into account, does not predict, in the high frequency region, for the real and imaginary parts of the impedance, the trends predicted by Warburg's impedance in the Nernstian approximation. The presented model can be generalized to the case of asymmetric cell, assuming boundary conditions physically sound.
\end{abstract}
 \maketitle

\section{Introduction}
The electric impedance of a linear system is defined as the ratio between the applied voltage and the current passing in the system itself. When the applied voltage is a simple harmonic function of the time, the impedance is a complex number, whose real part is related to the ohmic resistance of the system, and the imaginary part to its capacitive and inductive properties. When the system is an isotropic liquid containing ions, the impedance of the cell can be determined by means of a model based on the equations of continuity for the positive and negative ions, and on the equation of Poisson for the actual electric potential across the cell. This model has been called Poisson-Nernst-Planck (PNP) model, and discussed long ago by Macdonald \cite{macdonald}. In its simplest version the cell is a slab limited by blocking electrodes, the liquid contains impurities completely dissociated, and the ions are identical in all the aspects, but with charge equal in modulus and opposite in sign \cite{anca}. In this framework, the evaluated impedance is equivalent to that deduced for a medium described by a Debye's model for the dielectric constant with one relaxation time \cite{coelho}. In particular, the parametric plot of the imaginary part versus the real one is a semicircle, whose center is on the real axis, passing for the origin of the cartesian reference frame.
Several generalizations of the original model have been proposed to take into account the difference between the ions \cite{macdonald}, the electrodes properties \cite{m2,bata} and  the generation/ recombination of ions \cite{m3, lenzi, ioannis}, fractional diffusion or  boundary conditions taking into account memory effects \cite{lenzi2,lenzi3,lenzi4,lenzi5,lenzi6,lenzi7}.

An extension of the PNP model to evaluate the impedance of an insulating medium in which are injected ions is possible, when the mechanism describing the injection of ions is known. Several papers have been published on this subject for its technological importance \cite{us,kn}. In this type of papers, assuming an increasing of the charges carriers due to the difference of potential applied to the cell, and taking into account the trapping of the charge carriers in traps present in the medium, the impedance of the cell is determined for different types of electrodes \cite{ge,bi,bo,li,ju,moya}. According to the analyises presented in \cite{ge,bi,bo,li,ju,moya}, the parametric plot of the imaginary part versus the real part in the high frequency region is a straight line of slope one with respect to the real axis.

The analysis presented, among others, in \cite{ge,bi,bo,li,ju,moya} is based on three assumptions: 1) the increasing of the charge carriers close to electrodes is in agreement with Nernst equation, 2) the drift current is negligible with respect to the diffusion one, and 3) the displacement current is neglected in the evaluation of the impedance of the cell. These assumptions are questionable. In fact, Nernst equation gives the ionic concentration across to a membrane in a state of equilibrium, in the absence of a current. When a charge current is present it is a function of the over potential of the electrode with respect to the solution in contact with, as in Butler Volmer equation \cite{atkins}. Furthermore, the assumption that the drift current is negligible with respect to the diffusion one has to be checked at the end of the calculation, but this control has never been done in the quoted papers. In fact, if in the bulk the electric field, in the small voltage limit, is practically zero, this is not true close to the electrodes, where this term could be important. However, the third assumption, according to which the displacement current can be neglected is fatal in the analysis presented in \cite{ge,bi,bo,li,ju,moya} and in some text-books, as \cite{bockris}. In fact, as it is well known, the conduction current density in a one-dimensional system is not position independent, whereas the total current density, defined as the sum of the conduction and of the displacement current is constant across the cell \cite{m4}. It follows that in the framework of the analysis presented in \cite{ge,bi,bo,li,ju,moya} it is no clear why the current density has to be evaluated on one electrode, and not on the other. This is due to the fact that in the quoted analyses no mention to the Poisson's equation has been done, and the electric field is assumed to be negligible everywhere in the sample.

The aim of our paper is to show that when the analysis is correctly performed, taking into account the proper electric current density, even in the Nernstian approximation, the parametric plot of the imaginary part versus the real one does not present, in the high frequency region, the linear part. Hence, if this part of the parametric plot is experimentally observed, its explanation has to be searched in the injection mechanism.

\section{The physical problem}
Let us consider a sample in the shape of a slab of thickness $d$. The Cartesian reference frame has the $z$-axis perpendicular to the limiting surfaces, placed at $z=\pm d/2$. The problem is considered one-dimensional, in the sense that all the physical quantities depend just on the spatial coordinate $z$ and time $t$. The medium in between the electrodes  is considered, in the absence of an applied difference of potential, free of ions. Its dielectric constant is indicated by $\varepsilon$ and assumed to be frequency independent in the considered frequency range. When a difference of potential is applied to the sample, a quantity of particles of electric charge $q$ are injected into the sample. The diffusion coefficient of the charges in the medium is indicated by $D$ and their mobility by $\mu$. We assume the validity of Einstein's relation $\mu/D=q/(K_BT)$, where $K_BT$ is the thermal energy. In this situation, the evolution of the electric charge density and of the electric potential across the sample are well described by the continuity equation for  the diffusing particles and the equation of Poisson relating the actual potential to the bulk density of charges. We suppose that when the charges $q$ moves across the medium they can be trapped, and remain fixed in the trapping point. From this hypothesis it follows that we have two type of charges in the medium: the mobile and the fixed.  The first kind of charges contribute to the flux of electric charge and to the electric potential,$V$, the second type of charges contribute only to the actual potential profile. We indicate the corresponding bulk densities of charges by $n_m$ and $n_f$. The total bulk density of charges in the medium is $n=n_m+n_f$. The bulk density of electric charges in the medium, $j$, reduces to that of the mobile ones, $j_m$, and it is given by
\begin{equation}
\label{1}j=j_m=-D n_{m,z}+\mu n_m E.
\end{equation}
where we use the comma notation for the partial derivative $n_{m,z}=\partial n_m/\partial z$, and $E=- V_{,z}$ is the electric field. The continuity equations for the mobile and fixed charges, in the presence of trapping, are
\begin{eqnarray}
\label{2}n_{m,t}&=&- j_{m,z}+{\cal S},\\
\label{3}n_{f,t}&=&-{\cal S},
\end{eqnarray}
where ${\cal S}$ is a term taking into account the trapping phenomenon of the mobile charges. It represents a source for the fixed charges, and a well for the mobile ones. Taking into account (\ref{1}), Eq.(\ref{2}) can be rewritten as
\begin{equation}
\label{4}n_{m,t}=D\left\{ n_{m,z}+(q n_m/K_BT)\,V_{,z}\right\}_{,z}+{\cal S}
\end{equation}
In the following we limit our analysis to the case where the drift component of the bulk current density is negligible with respect to the diffusion one, i.e. 
\begin{equation}
\label{5}n_{m,z} \gg (q n_m/K_BT)\,V_{,z},
\end{equation}
that refers to a case of small applied potential to an insulating material \cite{ge,bi,moya}. The source term is assumed in the form ${\cal S}=-\kappa n_m$ as proposed by \cite{cusserl, crank}, and considered in details by \cite{ge,bi,bo,li,ju,moya}. This phenomenological term simply states that the bulk density of trapped charges is proportional to the bulk density of mobile charges, which is rather reasonable. In this framework the fundamental equations of the problem are
\begin{eqnarray}
\label{7} n_{m,t}&=&D\,n_{m,zz}-\kappa n_m,\\
\label{8}n_{f,t}&=&\kappa n_m,\\
\label{9}u_{v,zz}&=&-G(n_m+m_f),
\end{eqnarray}
where $u_v=q V/(K_BT)=V/V_{\rm th}$
is the electric potential expressed in thermal voltage units $V_{\rm th}=K_BT/q$. In Eq.(\ref{9}) $G=q/(\varepsilon V_{\rm th})$  is an intrinsic length of the problem.
The fundamental equations of the problem, (\ref{7},\ref{8},\ref{9}), are linear. Consequently if the applied voltage is a simple periodic function of the type $\Delta V(t)=V_0\,\exp(i \omega t)$, of amplitude $V_0$ and circular frequency $\omega$, the functions defining the dynamical state, $n_m(z,t)$, $n_f(z,t)$ and $u_v(z,t)$ are of the type
\begin{equation}
\label{12}[n_m,n_f,u_v]=[\phi_m(z),\phi_f(z),\phi_v(z)]\,\exp(i \omega t).
\end{equation}
Substituting the ansatz (\ref{12}) into (\ref{7},\ref{8},\ref{9}) we get
\begin{eqnarray}
\label{13}i \omega \phi_m&=&D \phi_m''-\kappa \phi_m,\\
\label{14}i \omega \phi_f&=&\kappa \phi_m,\\
\label{15}\phi_v''&=&-G(\phi_m+\phi_f),
\end{eqnarray}
where $f'=df/dz$. Equation (\ref{15}) shows that the fixed charges contribute to the effective potential across the sample, although they do not contribute to the bulk density current. Solution of the ordinary differential Eq.(\ref{13}) is
\begin{equation}
\label{16}\phi_m(z)=C_1\,\sinh(\beta z)+C_2\,\cosh(\beta z),
\end{equation}
where $C_1$ and $C_2$ are two integration constants to be determined by the boundary conditions, and $\beta=\sqrt{(\kappa+i \omega)/D}$  a complex wave number.
It depends on the reaction term $\kappa$ and on the circular frequency of the external voltage. The quantities
$\ell_{\kappa}=\sqrt{D/\kappa}$ and $\ell_{\omega}=\sqrt{\omega/D}$,
are two lengths, one related to the reaction term, the other to the diffusion. Solving Eq.(\ref{14}) the bulk density of fixed charges amplitude $\phi_f(z)$ is
\begin{equation}
\label{19}\phi_f=-i\frac{\kappa}{\omega}\,\phi_m,
\end{equation}
and the electric potential amplitude $\phi_v(z)$ is
\begin{equation}
\label{20}\phi_v(z)=i\frac{G D}{\omega}\,\phi_m(z)+C_3 z+ C_4.
\end{equation}
The integration constants $C_3$ and $C_4$ have to be determined, as $C_1$ and $C_2$, by the boundary conditions of the problem.

The total electric current density in the cell is given by
\begin{equation}
\label{21}j=q j_m+\varepsilon\, E_{,t}.
\end{equation}
It is constant across the cell \cite{m4}. In fact from (\ref{21}) it follows that
\begin{equation}
\label{22}j_{,z}=q\,j_{m,z}+\varepsilon E_{,zt},
\end{equation}
that is identically zero for the equations of continuity (\ref{2},\ref{3}). In the case under consideration taking into account (\ref{12}) we get
\begin{equation}
\label{23}j=-\varepsilon V_{\rm th} \left(G D \phi_m'+i \omega \phi_v'\right) e^{i \omega t}.
\end{equation}
Substituting (\ref{16}) and (\ref{20}) into (\ref{23}) we obtain, finally
\begin{equation}
\label{24}j=- i \omega \varepsilon C_3 V_{\rm th}\,e^{i \omega t},
\end{equation}
which is $z$-independent, as expected. The impedance of the cell, $Z=V_0/I$, with $I=j S$, where $S$ is the surface area of the electrodes, is then
\begin{equation}
\label{25}Z=-i \frac{u_0}{\omega \varepsilon C_3 S},
\end{equation}
where $u_0=V_0/V_{\rm th}$. The analysis presented above is general. It is valid for all boundary conditions. The boundary conditions on the electric potential are
\begin{equation}
\label{26} V(\pm d/2,t)=\pm (V_0/2)\,e^{i \omega t},
\end{equation}
stating that the electric potential of the electrodes coincides with that imposed by the external power supply.

\section{Application to a symmetric sample}

As a simple example let us consider the case where the two electrodes are the same. We suppose furthermore that the presence of the external electric potential is responsible for a density of charges on the electrode, that in the small voltage approximation is described by
\begin{equation}
\label{27}n(\pm d/2,t)=h V(\pm d/2,t),
\end{equation}
where $h$ is a phenomenological parameter describing the increase of the bulk density at the surface due to the external power supply. In the SI its units are 1/(m$^3$ V). In the case under consideration, for the analysis presented above, Eq.s(\ref{27}) are equivalent to
\begin{equation}
\label{28}\phi_m+\phi_f=\pm H\frac{u_0}{2},
\end{equation}
where $H=h V_{\rm th}$, for $z=\pm d/2$, respectively. The parameter $H$ has the dimension of the inverse of a volume. Due to the symmetry of the problem $\phi_m$, $\phi_f$ and $\phi_v$ are expected to be odd functions of $z$. It follows that in (\ref{16}) $C_2=0$, and in (\ref{20}) $C_4=0$. Using (\ref{26}) and (\ref{27}) we get for the other integration constants the expressions
\begin{eqnarray}
\label{29}C_1&=&i\frac{\omega_r}{\kappa_r+i \omega_r}\,\,\frac{H}{\sinh(\beta d/2)}\,\frac{u_0}{2},\\
\label{30}C_3&=&\frac{1+\kappa_r+i \omega_r}{\kappa_r+i \omega_r}\,\,\frac{u_0}{d},
\end{eqnarray}
where $\kappa_r=\kappa/\omega_0$, $\omega_r=\omega/\omega_0$, $\omega_0=DGH$, and in terms of reduced quantities $\beta=\sqrt{G H (\kappa_r+i\omega_r)}$. Note that in the framework of the present model $(GH)^{-1/2}$ is an intrinsic length of the problem related to the injection mechanism.
Substituting (\ref{30}) into (\ref{25}), and taking into account the definition of $\beta$, we obtain
\begin{equation}
\label{31}Z=R_0\,\frac{1-i\kappa_r/\omega_r}{1+ \kappa_r+i\omega_r},
\end{equation}
where $R_0=d/(\varepsilon \omega_0 S)$.
The characteristics frequency $\omega_0$ plays in the present problem the same role of Debye's circular frequency in the electric response of an electrolytic cell to an external electric field \cite{anca}. The real and imaginary parts of the electric impedance given by (\ref{31}) are
\begin{eqnarray}
\label{33}R=R_0\,\frac{1}{(1+\kappa_r)^2+\omega_r^2},\\
\label{34}X=-R_0\,\frac{\kappa_r(1+\kappa_r)+\omega_r^2}{(1+\kappa_r)^2+\omega_r^w}.
\end{eqnarray}
From Eq.(\ref{34}) it follows that the reactance of the cell  has, for $\kappa_r\leq \kappa^*=1/8$, two extrema for
\begin{equation}
\label{34-1}\omega_r=\omega_0\sqrt{\frac{1-\kappa_r(1+2\kappa_r)\pm(1+\kappa_r)\sqrt{1-8\,\kappa_r}}{2}}.
\end{equation}
On the contrary, for $\kappa_r>1/8$, the reactance is a monotonic function.
From (\ref{33}) and (\ref{34}) it is evident that for $\omega\to 0$
\begin{equation}
\label{35}R\to R_0\,\left(\frac{1}{1+\kappa_r}\right)^2,\quad X\to-R_0\,\frac{\kappa_r } {\omega_r(1+\kappa_r)},
\end{equation}
whereas in the opposite limit of $\omega\to \infty$
\begin{eqnarray}
\label{37}R\to R_0\,\omega_r^{-2},\quad
X\to -R_0\,\omega_r^{-1}.
\end{eqnarray}
From (\ref{35}) we get that in the dc limit $R$ tends to a constant value, whereas $X$ diverges as $1/\omega$. In the opposite limit  of very high frequency $R$ tends to zero as $1/\omega^2$, and $X$ as $1/\omega$. In this limit the correct  calculation does not predict Warburg's dependence for $R$ and $X$, contrary to the statement reported in many papers \cite{ge,bi,bo,li,ju,moya} and text-book \cite{bockris}.

For $\kappa_r=0$ the parametric plot of $X$ versus $R$ is a semicircle. The parametric plot  for $\kappa_r\neq 0$ has a vertical asymptote defined by Eqs.s(\ref{35}). Increasing $\kappa_r$ the asymptote moves to the left.

\section{Limit of the approximations}
We have now to analyze when the condition $n_{m,z}\gg n_m\ u_{v,z}$, in which the drift component of the current density is negligible with respect to the diffusion one is verified. Taking into account the analysis presented above this condition can be rewritten as
\begin{equation}
\label{38}|\phi_m'|\gg |\phi_m \phi_v'|.
\end{equation}
If it is verified for $z=\pm d/2$ it is also verified everywhere. Furthermore, it depends on the frequency. For the case under consideration a simple calculation gives, in the limit for $\omega_r\to 0$ at $z=d/2$,
\begin{eqnarray}
\phi_m'&=&H\frac{\sqrt{G H \kappa_r}\,\coth(\sqrt{G H \kappa_r}\,d/2)}{2 \kappa_r}u_0 \omega_r,\nonumber\\
\phi_m\phi_v'&=&H\frac{2(1+\kappa_r)-d\sqrt{G H\kappa_r}\coth(\sqrt{G H \kappa_r}d/2)}{4 d \kappa-r^2}\,u_0^2 \omega_r.\nonumber
\end{eqnarray}
If $\sqrt{G H \kappa_r} d/2\gg 1$ these relations become
\begin{eqnarray}
\phi_m'&=&H\frac{\sqrt{G H \kappa_r}}{2 \kappa_r}u_0 \omega_r,\nonumber\\
\phi_m\phi_v'&=&H\frac{2(1+\kappa_r)-d\sqrt{G H\kappa_r}}{4 d \kappa-r^2}\,u_0^2 \omega_r,\nonumber
\end{eqnarray}
and the condition $|\phi_m'|\gg |\phi_m \phi_v'|$ gives
\begin{equation}
\label{39}G H >>\frac{1}{\kappa_r}\,\left(2\,\frac{(1+\kappa_r) u_0}{(2\kappa_r+u-0)d}\right).
\end{equation}
In the opposite limit where $\sqrt{G H \kappa_r} d/2\ll 1$, when $\omega \to 0$ and for $z=d/2$ we get
\begin{equation}
\phi_m'=\frac{H u_0}{d \kappa_r}\,\omega_r,\quad \phi_m \phi_v'=\frac{H u_0^2}{2 d \kappa_r}\,\omega_r,\nonumber
\end{equation}
and the condition $|\phi_m'|\gg |\phi_m \phi_v'|$  implies that $u_0\gg 2 $. Since the analysis is valid only for small $u_0$ we conclude that in the present model has to be verified the condition $\sqrt{G H \kappa_r} d/2\gg 1$. In the opposite limit where $\omega_r\to \infty$ we obtain, at $z=d/2$,
\begin{equation}
\phi_m'=\frac{H \sqrt{G H}}{2}\,\sqrt{\omega_r},\quad \phi_m \phi_v'=\frac{H}{2 d} u_0^2.
\end{equation}
It follows that in the high frequency region the drift current is always negligible with respect to the diffusion one, as assumed in \cite{ge,bi,bo,li,ju,moya}.

Let us consider, finally, the importance of the displacement current with respect to the conduction current, due to diffusion. The total current is give by (\ref{21}), that with the present symbols reads
\begin{equation}
\label{40}j=-D\left(\phi_m'+i H \omega_r \phi_v'\right)\,\exp(i \omega t).
\end{equation}
since the electric field is localized close to the electrodes, we compare the terms $\phi_m'$ and $H \omega_r \phi_v'$ for $z=\pm d/2$. Using for $\phi_m$ and $\phi_v$ the expression reported above, in the limit $\sqrt{G H \kappa_r} d/2\gg 1$ we get
\begin{equation}
\label{41}|\phi_m'|=|i H \omega_r \phi_v'|=\frac{DH\omega_r}{2}\,\sqrt{\frac{G H}{\kappa_r+i \omega_r}}\,u_0.
\end{equation}
This result indicate that the displacement current  can never be neglected in the analysis.

\section{Conclusions}
We have determined the electric impedance of a cell made by an insulating medium where, the charge carriers are injected by means of an external power supply. The presence of trap, described by a first order chemical reaction, has been taken into account. In the framework where the conduction current is due just to diffusion, we have shown that, in the high frequency range, the correct expression of the impedance does not have Warburg's characteristics. Previous analyses published by several authors have shown a linear dependence in the high frequency region, reactance versus the resistance of the impedance, in the series representation. However, since these analyses are based on the assumption that the displacement current can be neglected with respect to the diffusion one, they are not correct. We show on a simple case very often considered in literature, where the bulk density of charge carriers at the surface is a function of the applied potential, that these two currents are equal. From the analysis reported in our communication it follows that if a Warburg type impedance is observed in the high frequency region, its origin is not due to a simple mechanism of diffusion. It can be related to injection mechanism, according to which the injected current does not depend only on the surface potential. The presented model can be generalized to more realistic boundary conditions.


\begin{thebibliography}{99}
\bibitem{macdonald}J. Ross Macdonald, Phys. Rev. \textbf{92}, 4 (1953).
\bibitem{anca}G. Barbero, A. L.  Alexe-Ionescu, Liquid Crystals, {\bf 32}, 943 (2005).
\bibitem{coelho}R. Coelho, "\textit{Physics of Dielectrics for Engineer}",
Elsevier Scientific Publishing Company, Amsterdam, 1979.
\bibitem{m2}J. Ross Macdonald, D. Franceschetti, Electroanal. Chem. {\bf 82}, 271 (1977).
\bibitem{bata}G. Barbero, F. Batalioto, A. M. Figueiredo Neto, J. Appl.
Phys. \textbf{101}, 054102 (2007).
\bibitem{m3}J. Ross Macdonald, D. Franceschetti, J. Chem. Phys. {\bf 68}, 1614 (1978).
\bibitem{lenzi}G. Derfel, E. Kaminski Lenzi, C. Refosco Yednak, G. Barbero, J. Chem. Phys. {\bf 132}, 224901 (2010).
\bibitem{ioannis}I. Lelidis, G. Barbero, and A. Sfarna,
J. Chemical Physics. {\bf 137}, 154104 (2012).
\bibitem{lenzi2} E. K. Lenzi, P. R. G. Fernandes, T. Petrucci, H. Mukai, and
H. V. Ribeiro, Phys. Rev. E \textbf{84}, 041128 (2011).
\bibitem{lenzi3} P. A. Santoro, J. L. de Paula, E. K. Lenzi, and L. R.
Evangelista, J. Chem. Phys. \textbf{135}, 114704 (2011).
\bibitem{lenzi4} J. L. de Paula, P. A. Santoro, R. S. Zola, E. K. Lenzi, L.
R. Evangelista, F. Ciuchi, A. Mazzulla, and N. Scaramuzza, Phys. Rev. E
\textbf{86}, 051705 (2012).
\bibitem{lenzi5} F. Ciuchi, A. Mazzulla, N. Scaramuzza, E. K. Lenzi, and L.
R. Evangelista, J. Phys. Chem. C \textbf{116}, 8773 (2012).
\bibitem{lenzi6} E. K. Lenzi, P. R. G. Fernandes, H. Mukai, H. V. Ribeiro,
M. K. Lenzi, and G. Goncales, Int. J. Electrochem. Sci. \textbf{8}, 2849
(2013).
\bibitem{lenzi7} E. K. Lenzi, J. L. de Paula, F. R. G. B. da Silva, L. R.
Evangelista, J. Phys. Chem. C \textbf{117}, 23685 (2013).
\bibitem{us} M. Usman Iftiktar, D. Riu, F. Druat, S. Rosini, Y. Bultel, and N. Retiere, J. Power Sources, {\bf 160}, 1170 (2006).
\bibitem{kn}E. Kniaginicheva, N. Pismenskaya, S. Melikinov, E. Belashova, P. Sistat, M. Cretin, and V. Nikonenko, J. Membr. Sci. {\bf 496}, 78 (2015).
\bibitem{ge}H. Gerischer, Z. Phys. Chem. {\bf 198}, 266 (1951).
\bibitem{bi}J. Bisquert, J. Phys. Chem. {\bf 106}, 325 (2002).
\bibitem{bo}B. A. Boukamp, H. J. M. Bouwmeester, Solid State Ionics, {\bf 157}, 29 (2003).
\bibitem{li} F. Li, J. R. Jennings, Q. Wang, J. Chua, N. Mathews, S. G. Mhaisalkar, S-J. Moon, S. M. Zakeeruddin, and M. Graetzel, J. Phys. Chem. C, {bf 117}, 10980 (2013).
\bibitem{ju} R. Jurczakowski and P. Polczynski, J. Phys. Chem. C, {\bf 118}, 7980 (2014).
\bibitem{moya}A. A. Moya, Phys. Chem. Chem. Phys. {\bf 18}, 3812 (2016).
\bibitem{atkins}P. W. Atkins, "{\it Physical Chemistry. Fifth Edition}", Oxford University Press, Oxford 1994.
\bibitem{bockris}J. O'M. Bockris and Amulya K. N. Reddy, "{\it Modern Electrochemistry}", Second Edition.
\bibitem{m4}J. Ross Macdonald, J. Appl. Phys. {\bf 46}, 4602 (1975).
\bibitem{cusserl}E. L. Cussler. "{\it Diffusion: Mass Transfer in Fluid System}" Cambridge University Press, Cam-
bridge, (1985).
\bibitem{crank}J. Crank, "{\it The Mathematics of Diffusion}", Oxford University Press, Oxford, 1965.


\end{thebibliography}
\end{document}